\documentclass[%
 reprint,
superscriptaddress,
showpacs,
 amsmath,amssymb,
prb,
]{revtex4-1}

\usepackage{epsfig}
\usepackage{xfrac}
\usepackage{siunitx}
\usepackage{graphicx}
\usepackage{dcolumn}
\usepackage{bm}

\usepackage{color}

\begin{document}


\title{Polarization memory in the nonpolar magnetic ground state of multiferroic CuFeO$_2$} 

\author{J. Beilsten-Edmands}
\email{james.beilsten-edmands@physics.ox.ac.uk}
\affiliation{\footnotesize Clarendon Laboratory, Department of Physics, University of Oxford, Parks Road, Oxford, OX1 3PU, United Kingdom}

\author{S. J. Magorrian}
\affiliation{\footnotesize Clarendon Laboratory, Department of Physics, University of Oxford, Parks Road, Oxford, OX1 3PU, United Kingdom}

\author{F. R. Foronda}
\affiliation{\footnotesize Clarendon Laboratory, Department of Physics, University of Oxford, Parks Road, Oxford, OX1 3PU, United Kingdom}

\author{D. Prabhakaran}
\affiliation{\footnotesize Clarendon Laboratory, Department of Physics, University of Oxford, Parks Road, Oxford, OX1 3PU, United Kingdom}

\author{P. G. Radaelli}
\affiliation{\footnotesize Clarendon Laboratory, Department of Physics, University of Oxford, Parks Road, Oxford, OX1 3PU, United Kingdom}

\author{R. D. Johnson}
\affiliation{\footnotesize Clarendon Laboratory, Department of Physics, University of Oxford, Parks Road, Oxford, OX1 3PU, United Kingdom}


\begin{abstract}
We investigate polarization memory effects in single-crystal CuFeO$_2$, which has a magnetically-induced ferroelectric phase at low temperatures and applied $\textit{B}$ fields between 7.5 and \SI{13}{\tesla}. 
Following electrical poling of the ferroelectric phase, we find that the nonpolar collinear antiferromagnetic ground state at $B =$  \SI{0}{\tesla}  retains a strong memory of the polarization magnitude and direction, such that upon re-entering the ferroelectric phase a net polarization of comparable magnitude to the initial polarization is recovered in the absence of external bias. This memory effect is very robust: in pulsed-magnetic-field measurements, several pulses into the ferroelectric phase with reverse bias are required to switch the polarization direction, with significant switching only seen after the system is driven out of the ferroelectric phase and ground state either magnetically (by application of $B >$ \SI{13}{\tesla}) or thermally.  The memory effect is also largely insensitive to the magnetoelastic domain composition, since no change in the memory effect is observed for a sample driven into a single-domain state by application of stress in the [1$\overline{1}$0] direction.  On the basis of Monte Carlo simulations of the ground state spin configurations, we propose that the memory effect is due to the existence of helical domain walls within the nonpolar collinear antiferromagnetic ground state, which would retain the helicity of the polar phase for certain magnetothermal histories.
\end{abstract}

\maketitle

\section{Introduction}

The discovery of type-II multiferroics, where ferroelectricity is induced by magnetic order, has generated significant interest over the past decade.
Among the large family of compounds now known to display this phenomenology, the CuFeO$_2$ delafossite is particularly interesting because ferroelectricity induced by an incommensurate proper-screw magnetic order, resulting in a handedness-dependent polarization, was first discovered in this material \cite{Mitsuda2000, Kimura06, NakajimaJPSJ07, NakajimaPRB08}.  Neither of the multiferroic mechanisms known until then (spin-current/cycloidal and exchange striction) was found to be applicable to CuFeO$_2$; later, Arima proposed a new mechanism, known as spin-dependent $\textit{d-p}$ hybridization \cite{Arima} to explain the multiferroic properties of this material.

The magnetic and structural phase diagram of CuFeO$_2$ has been extensively characterized, and is now fairly well established. In the absence of a magnetic field, CuFeO$_2$ undergoes two magnetic transitions upon cooling, which are accompanied by structural distortions due to magnetoelastic coupling \cite{Terada06, YePRL06, Teradajpsj06}. Above $T_{\text{N1}}$ = \SI{14}{\kelvin}, CuFeO$_2$ is paramagnetic and its crystal structure is described by the $R\overline{3}m$ space group. At $T_{\text{N1}}$, the Fe$^{\text{3+}}$ spins magnetically order into a collinear-incommensurate state [spin-density wave (SDW)]\cite{Mitsuda98}, with wave vector $Q =$ ($q_{\text{IC}},q_{\text{IC}}$,$\sfrac{3}{2}$), where $0.19<q_{\text{IC}}(T)<0.22$ . At $T_{\text{N2}}$ = \SI{11}{\kelvin}, the spins lock into a collinear constant-moment $\uparrow \uparrow \downarrow \downarrow$ four-sublattice (4SL) ground state, with $Q =$ ($\sfrac{1}{4}$,$\sfrac{1}{4}$,$\sfrac{3}{2}$). 
Magnetoelastic distortions lower the crystal symmetry to $\it{C}$2/$\it{m}$ at $T_{\text{N1}}$ and result in a scalene-triangle distortion at $T_{\text{N2}}$, which lifts the degeneracy of the in-plane exchange interactions to stabilise the 4SL structure \cite{Terada06, Teradajpsj06}. 
The second transition is associated with the loss of $C$ centering, as confirmed by the the appearance of additional Bragg reflections \cite{Teradajpsj06,Terada08}, further lowering the symmetry to either $P$2/$c$ or $P$2, depending on whether or not inversion symmetry is maintained.   
Upon application of a magnetic field $B$ along the trigonal $c$ axis, CuFeO$_2$ undergoes a further series of magnetic transitions \cite{Lummen09}. Between 7.5 and \SI{13}{\tesla}, the $\text{Fe}^{3+}$ spins form an incommensurate proper-screw magnetic structure, which we refer to as the ferroelectric incommensurate (FEIC) phase, with $Q =$ ($q_{\text{IC}},q_{\text{IC}}$,$\sfrac{3}{2}$) where $q_{\text{IC}} \sim 0.21$. This produces a ferroelectric polarization along the [110] direction, with the sign of $\textit{P}_{[110]}$ being determined by the handedness of the magnetic structure. At \SI{13}{\tesla} there is a transition to a $\uparrow \uparrow \uparrow \downarrow \downarrow$ five-sublattice (5SL) phase with $Q =$ ($\sfrac{1}{5}$,$\sfrac{1}{5}$,$\sfrac{3}{2}$) that is accompanied by spin-driven bond ordering which removes the scalene-triangle distortion and restores the crystal symmetry to $\it{C}$2/$\it{m}$ \cite{NakajimaPRB13}. There are further magnetic transitions at 20 and \SI{33}{\tesla} to a $\uparrow \uparrow \downarrow$ three-sublattice (3SL) and canted-3SL phase, respectively.

By contrast, the dielectric and ferroelectric properties of CuFeO$_2$ are still the subject of significant debate.  A \emph{history dependence} of the ferroelectric polarization of CuFeO$_2$ was measured in pulsed $B$ fields by Mitamura \textit{et al.} \cite{Mitamura07}, whereby several $B$-field pulses are needed to saturate the value of $P$. This was suggested to result from successive domain alignment and repopulation of the three magnetoelastic $q$ domains that are related by the broken threefold symmetry, but its origin is far from clear.  Another issue that remains to be clarified is the observation in pulsed-field measurements of a \emph{residual polarization} in the 4SL and 5SL phases (i.e. the measured polarization does not return to zero upon leaving the FEIC phase) \cite{Mitamura07, Mitamura06}.
Furthermore, a \emph{ferroelectric memory effect} has been reported for the analogous phase in CuFe$_{1-x}$Ga$_x$O$_2$ \cite{Mitsuda2009,NakajimaJPSJ13}, 
whereby a net ferroelectric polarization is recovered after warming out of the ferroelectric phase and then cooling without electrical bias, which was attributed to residual charges trapped at magnetoelastic $q$-domain boundaries. A polarization memory effect occurs when the direction and strength of the electrical polarization are retained through a history involving a non-ferroelectric phase, achieved through either temperature or magnetic field cycling.

In this paper, we present a series of  pyrocurrent and magnetocurrent measurements on conventional and mechanically detwinned (single-$q$-domain) single crystals, which, we believe, completely clarify these issues.  The residual polarization in zero magnetic field is found to be an artifact of the pulsed-field measurements, in which $\textit{P}_{[110]}$ is typically measured in an applied electric field bias, leading to magnetic-field-dependent leakage currents.  We determine that no residual polarization exists when measurements are performed in zero bias and conclude that the 4SL phase is centrosymmetric (space group $P$2/$c$) within our sensitivity.  By contrast,  we confirm the existence of a strong polarization memory effect, which persists in the \emph{nonpolar} 4SL phase of CuFeO$_2$. We also establish that this memory effect is not related to magnetoelastic $q$ domains and postulate the existence of helical domain walls between antiferromagnetic phase domains --- a hypothesis that would fully explain the observed memory effects and that is supported by Monte Carlo simulations of the ground state.

\section{Experimental Procedure}

\begin{figure}
\includegraphics[width=0.5\textwidth]{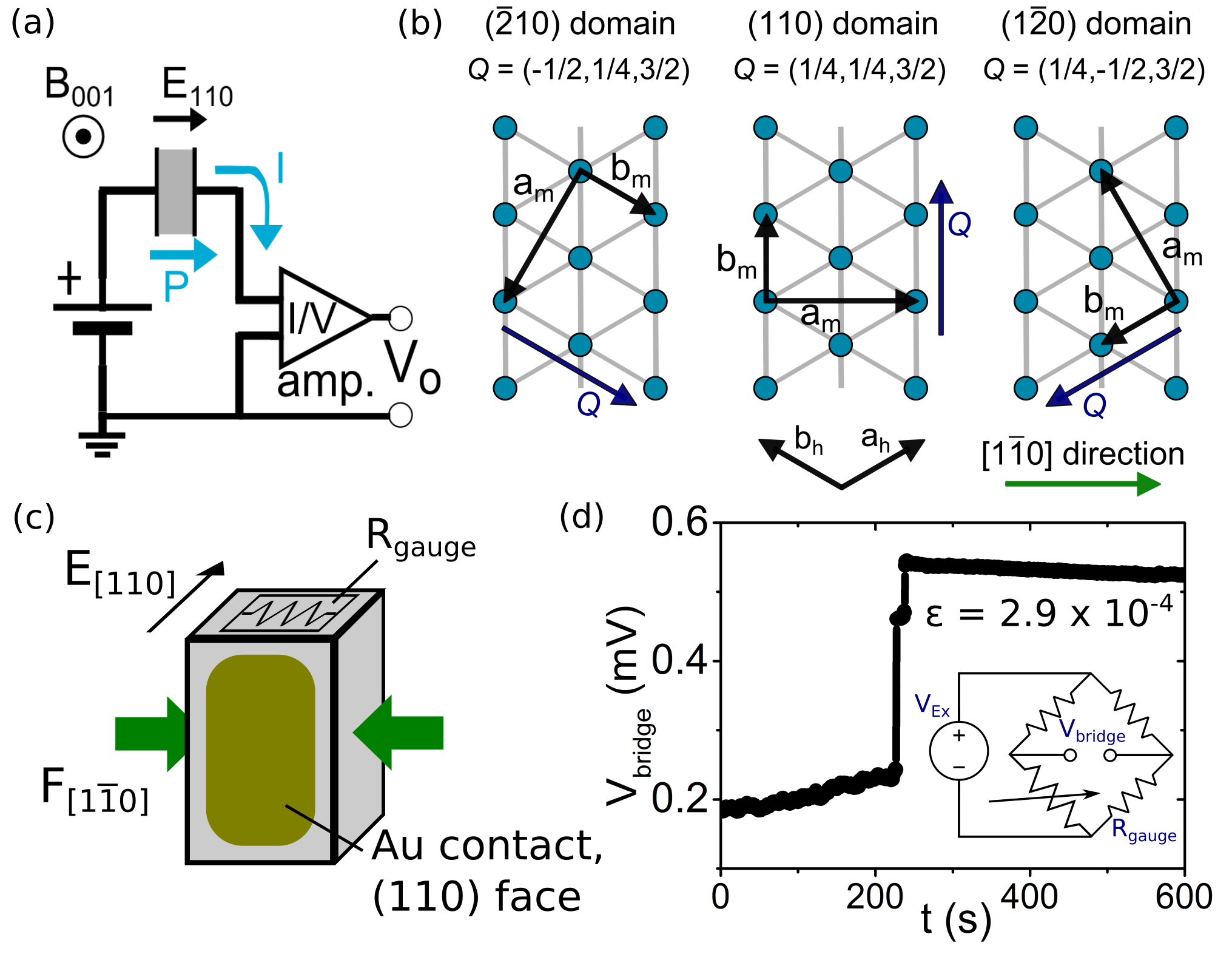}
\caption{(a) Circuit diagram for pyrocurrent/magnetocurrent measurements using a transimpedance amplifier. In pulsed-field measurements, the voltage source provides an $E$ field (the ``bias'' field) during the measurement, which only lasts \SI{14}{\milli\second}. In persistent $B$-field measurements, the ``poling'' $E$ field  is only applied upon cooling and then removed (during the measurement of $P$, the left hand side electrode is connected to ground). (b) Relationship between the axes of the three monoclinic domains and the parent hexagonal structure. The monoclinic domains are distorted relative to the hexagonal structure, with an increase in $b_m$ and a decrease in $a_m$. (c) For one set of measurements, uniaxial force was applied to the crystal along the [1$\overline{1}$0] direction at room temperature in order to favor the $Q$ = (q,q,3/2) domain upon cooling through $T_{\text{N1}}$. The strain on the sample was measured using a strain gauge (R$_\text{gauge}$) mounted on top of the sample and a balanced bridge circuit. (d) The strain was measured upon destraining (occuring from $t =$ \SI{220}{\second} to \SI{240}{\second}) to be $2.9 \times 10^{-4}$.}
\label{fig:exp}
\end{figure} 

Single crystals of CuFeO$_2$ were grown using a floating-zone method, as described in detail in the Appendix. Samples were structurally characterized using single-crystal x-ray diffraction, using a Rigaku Oxford Diffraction SuperNova diffractometer and Rigaku SmartLab X-ray diffractometer. The crystals were found to be essentially a single obverse/reverse domain, with less than 3\% fraction of the other twin present. 
Cuboid samples for polarization measurements were cut to give a large (110) face (area $\sim$\SI{15}{\square\milli\metre}), with a thickness of \SI{0.5}{\milli\metre} for unstrained samples and \SI{1.0}{\milli\metre} for strained samples (see below). X-ray micro-diffraction revealed a small degree of mosaicity, corresponding to the existence of large crystallites, with dimensions of the order of several millimeters, aligned along a common $c$ axis but with a slight misalignment, up to \SI{5}{\degree}, in the $ab$ plane. As such, the applied electric fields, strain and measured polarization will have components $>$99\% along the desired crystal directions of all crystallites, giving a negligible effect due to the mosaicity. Gold electrodes were evaporated onto the (110) faces and wires were contacted using silver paint.

The magnetization of the crystals was measured as a function of temperature using a Quantum Design MPMS;  we observed features at 11 and \SI{14}{\kelvin} in the dc magnetization curve at $B =$ \SI{1.0}{\tesla}, in agreement with previous reports \cite{Kimura06}.
Pulsed-field magnetization and polarization measurements were carried out up to 40 and \SI{18}{\tesla} respectively at the Oxford high-field magnet facility, using a pulse timewidth of \SI{14}{\milli\second}. 
A series of magnetic transitions and plateaus in magnetization was observed, in good agreement with previous reports \cite{Lummen09}.

In this paper we consistently apply the following terminology:  a \emph{biasing voltage} is maintained throughout the polarization measurements, whereas a \emph{poling voltage} is applied upon cooling and subsequently removed, following short-circuiting of the sample.
For polarization measurements in a pulsed field, samples were first cooled to \SI{4.2}{\kelvin} in zero field, before a \emph{biasing voltage} was applied and maintained throughout the duration of the \SI{14}{\milli\second} pulse. Polarization measurements in persistent magnetic fields were performed in an Oxford Instruments \SI{13.5}{\tesla} superconducting high field magnet using the following procedure. An external \emph{poling voltage} was applied to the samples upon cooling from \SI{15}{\kelvin} at \SI{10}{\tesla} using a Trek 610E high-voltage supply. Once at \SI{5}{\kelvin}, the poling field was removed and the sample electrodes were short-circuited for at least $10$ min. Some persistent-field measurements were also performed while applying a biasing voltage (see Fig. \ref{fig:memory} below).
In both pulsed and persistent fields, the ferroelectric polarization was determined by continuously integrating the current that flows to compensate the ferroelectric polarization density at the sample surface, as either the temperature or the magnetic field is varied: these are referred to as pyrocurrent and magnetocurrent measurements. The current was amplified using a Femto Transimpedance Amplifier and integrated analytically, after subtracting the background current determined from the current before and after the temperature or field sweep. 
The circuit diagram is shown in Fig. \ref{fig:exp}(a).

In order to investigate domain effects, the formation of a single magnetoelastic $q$ domain was induced by applying uniaxial stress to \SI{1.0}{\milli\metre} thick samples at room temperature using a custom-built device. 
The magnetic ordering below $T_{\text{N1}}$ breaks the threefold symmetry of the trigonal phase and magnetostriction gives rise to a monoclinic distortion. In the absence of stress, this results in three monoclinic domains related by the broken threefold symmetry of the trigonal phase, with a one-to-one correspondence between the magnetic wave vectors and the twofold monoclinic axes.
The three wave vectors in the 4SL phase are, in hexagonal coordinates,
($\sfrac{1}{4}$,$\sfrac{1}{4}$,$\sfrac{3}{2}$), ($\sfrac{1}{4}$,$-\sfrac{1}{2}$,$\sfrac{3}{2}$) and ($-\sfrac{1}{2}$,$\sfrac{1}{4}$,$\sfrac{3}{2}$), which we refer to as the (110),(1$\overline{2}$0) and ($\overline{2}$10) $q$-domains. The relationship between the monoclinic and the hexagonal settings for the three $q$ domains are shown in Fig.~\ref{fig:exp}(b).
Upon cooling through $T_{\text{N1}}$ the unit cell length is increased along the monoclinic $b$ axis ($\Delta b /b= 1.9 \times 10^{-3}$) and reduced along the monoclinic $a$ axis ($\Delta a /a = -1.3 \times 10^{-3}$ ) \cite{Teradajpsj06}. 
For the (110) domain $\Delta x_{[1\overline{1}0]} /x_{[1\overline{1}0]} = \Delta a /a <0$, whereas for the (1$\overline{2}$0) and ($\overline{2}$10) domains $\Delta x_{[1\overline{1}0]} /x_{[1\overline{1}0]}= \frac{1}{2}(\Delta a /a + \Delta b /b) > 0$. Therefore, applying uniaxial force along the [1$\overline{1}$0] direction favors the (110) crystallographic domain upon cooling through $T_{\text{N1}}$. 
The compressive strain along the [1$\overline{1}$0] direction at room temperature was measured upon destraining to be $2.9 \times 10^{-4}$ using a vishay strain gauge adhered to the top of the sample and a balanced bridge circuit. Based on the high-pressure lattice parameters reported in Ref. [\cite{Zhao97}], this corresponds to a pressure of approximately \SI{110}{\mega\pascal}.

\section{Results}
\subsection{Pulsed-field measurements}

\begin{figure}
\includegraphics[width=0.5\textwidth]{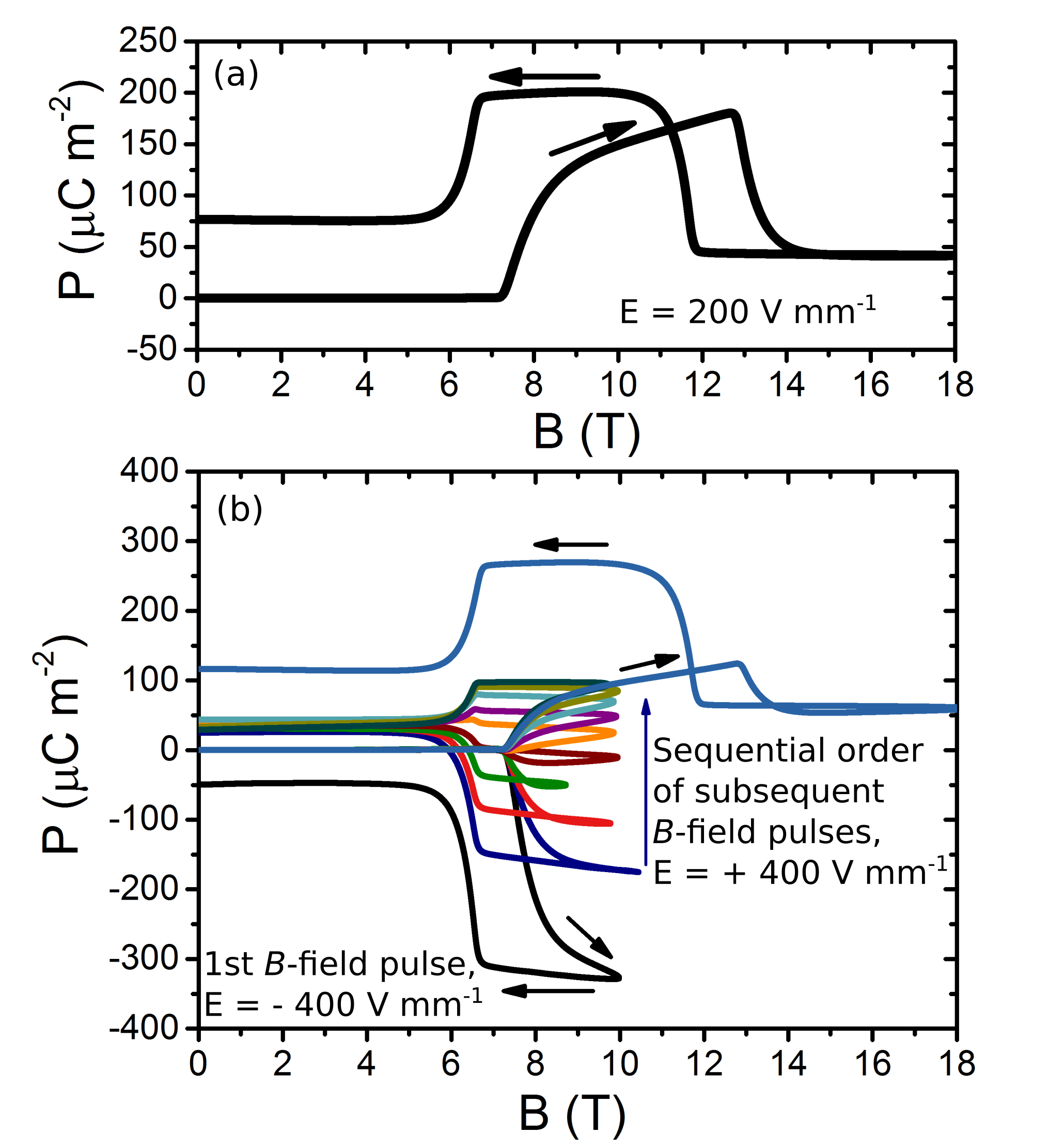}
\caption{(a) $B$-field dependence of polarization along [110] measured during an \SI{18}{\tesla} field pulse. 
The measured polarization in the 4SL and 5SL phase does not return to zero, giving the appearance of residual polarization. (b) A strong history dependence, or polarization memory, is observed for sequential measurements. After an initial \SI{10}{\tesla} $B$-field pulse with a negative electrical bias (\SI{-400}{\volt\per\milli\metre}), several pulses with a reversed bias (+\SI{400}{\volt\per\milli\metre}) are required to switch the direction of $P$. A significant polarization reversal is only obtained on the 11th pulse after driving the system into the 5SL phase.}
\label{fig:pf}
\end{figure}

The ferroelectric polarization of CuFeO$_2$ was first investigated in pulsed magnetic fields.
Figure~\ref{fig:pf}(a) shows $P_{[110]}$ for $B \parallel c$, measured with an applied biasing electric field $E$ of \SI{200}{\volt\per\milli\metre} and no applied stress. Upon increasing $B$, a polarization of \SI{180}{\micro\coulomb\per\square\metre} is measured in the FEIC phase between 7.5 and \SI{13}{\tesla}, however $P$ is not observed to return to zero above \SI{13}{\tesla} in the 5SL phase. Furthermore, upon returning from the FEIC phase to the 4SL phase by decreasing $B$, a residual polarization of \SI{75}{\micro\coulomb\per\square\metre} is measured at \SI{0}{\tesla}. These observations are consistent with a previous report of a residual polarization in both the 4SL and the 5SL phase \cite{Mitamura07}.

A strong memory effect was also observed for sequential pulsed-field measurements, as shown in Fig.~\ref{fig:pf}(b), where the zero of $P$ is redefined at the start of each measurement for clarity. 
First, $\textit{P}$ was measured for a \SI{10}{\tesla} pulse with $E$ = \SI{-400}{\volt\per\milli\metre}, giving a $\textit{P}$ of \SI{-310}{\micro\coulomb\per\square\metre}. In all subsequent measurements, a bias of +\SI{400}{\volt\per\milli\metre} was applied. During a transition from a non-ferroelectric to a ferroelectric phase with an applied bias, it is energetically favorable for $\textit{P}$ to align with $E$ and form a ferroelectric monodomain; therefore upon switching $E$ one would expect the direction of $\textit{P}$ to switch also.
In fact, upon a reverse in $E$ a polarization of \SI{-160}{\micro\coulomb\per\square\metre} was observed, and with each subsequent measurement $\textit{P}$ gradually increased in the direction of $E$, until the magnitude of $\textit{P}$ started to saturate at \SI{+100}{\micro\coulomb\per\square\metre}. A large, but not fully reversed, $\textit{P}$ of \SI{+270}{\micro\coulomb\per\square\metre} was only recovered once an \SI{18}{\tesla} pulse was applied to drive the system into the 5SL phase. 
In another set of measurements with $\left|E\right|$ = \SI{200}{\volt\per\milli\metre} and \SI{18}{\tesla} $B$-field pulses (not shown), it was found that two \SI{18}{\tesla} $B$-field pulses were needed before the direction of $\textit{P}$ was reversed.  This indicates that a memory of the polarization state is retained in both the 4SL and the 5SL phases, but that the memory appears to be stronger in the 4SL ground state.  The changing magnitude of $\textit{P}$ between subsequent measurements indicates that the population ratio of the two types of ferroelectric domains is gradually changed by each $B$-field pulse, as opposed to the complete switching from one domain type to another that would be expected.

Although the results reported here are broadly consistent with the literature, we wish to highlight a number of aspects that either have not been previously reported or have not been emphasised in the past.  First, we note that the residual polarization at $B$ = \SI{0}{\tesla} follows the direction of the \emph{bias} rather than the sign of the high-field ferroelectric polarization, and, for example, is similar for all \SI{10}{\tesla} pulses collected at $E$ = \SI{+400}{\volt\per\milli\metre} in spite of the fact that both the absolute values and the sign of the high-field ferroelectric polarization vary greatly between pulses. This strongly suggests that the residual polarization is due to a spurious current caused by the applied bias which gives a net non-zero charge when integrated over the whole pulse, rather than a physical ferroelectric polarization. This current is likely to be greater through the transitions, but it is impossible to determine exactly because of hysteresis. We discuss this further in light of our measurements in persistent magnetic fields. Secondly, although history dependences of the polarization in CuFeO$_2$ and memory effects in the Ga doped compound have been reported previously, this is, to the best of our knowledge, the first observation that the high-field polarization can be \emph{opposite} to the bias field.  This fact, together with the previous observation of significant leakage currents in biased measurements, rules out the possibility that the memory effect may be due to charge trapping either at the magnetoelastic domain boundaries \cite{Mitsuda2009,NakajimaJPSJ13}, at the mosaic crystallite boundaries (as observed by x-ray diffraction), or at the electrodes, since any trapped charges would be removed by the reversed bias (later, we show that the memory effect also exists in single-$q$-domain samples).  Another explanation for the memory effect is a small amount of phase coexistence of the FEIC phase in the 4SL and 5SL phases. Similar memory effects have been observed in other multiferroics such as CuO \cite{WuCuO2010} and MnWO$_4$ \cite{Taniguchi09}. In these cases, a memory of the polarity of an incommensurate multiferroic phase is retained in a collinear antiferromagnetic phase, which was attributed to multiferroic nanoregions. Such residual FEIC regions would act as nucleation centers upon reentering the FEIC phase and could therefore act to favor a magnetic structure of one handedness (and hence polarity) over the other.  We discuss this scenario in the remainder of the paper.

\subsection{Persistent-field measurements}

We further investigated the memory effects by performing measurements in persistent magnetic fields with zero bias, including measurements under applied strain, as described in Sec. II.

\begin{figure}
\includegraphics[width=0.5\textwidth]{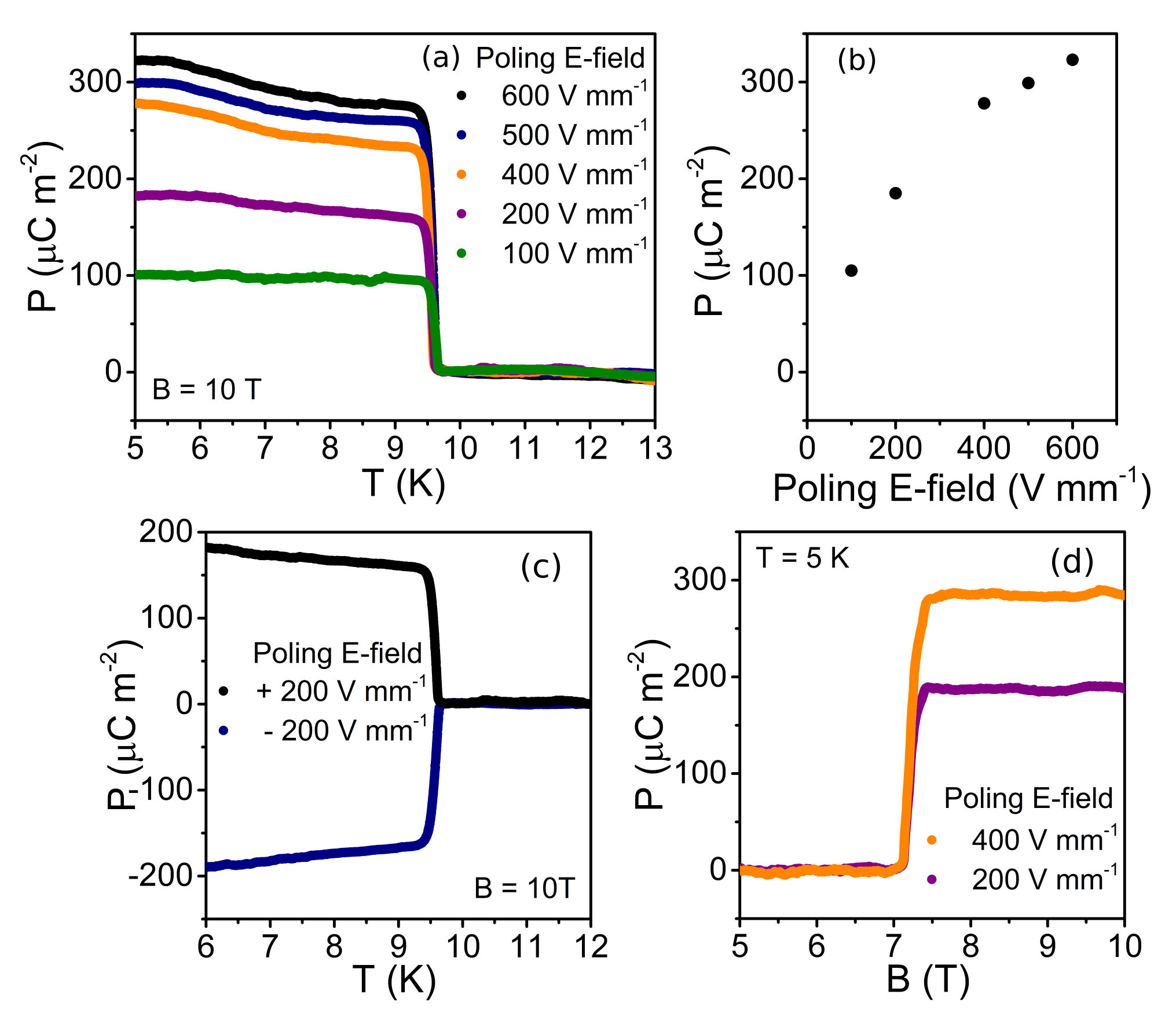}
\caption{(a), (b) Dependence of polarization on the poling $E$ field for persistent $B$-field measurements. The zero of polarization is defined as that at T = \SI{10}{\kelvin}. At T = \SI{5}{\kelvin}, saturation of $P$ $\sim$ \SI{325}{\micro\coulomb\per\square\metre} is only achieved at high poling $E$ fields above  $\sim$ \SI{500}{\volt\per\milli\metre}. (c) The direction of polarization is fully reversible by the application of a reversed poling $E$ field upon cooling, as expected for a ferroelectric. (d) Magnetocurrent measurements of the polarization, which give good agreement with the pyrocurrent measurements for the magnitude of $P$ and hence give no indication of a physical residual ferroelectric polarization in the 4SL phase.}
\label{fig:bias}
\end{figure}

\begin{figure}
\includegraphics[width=0.5\textwidth]{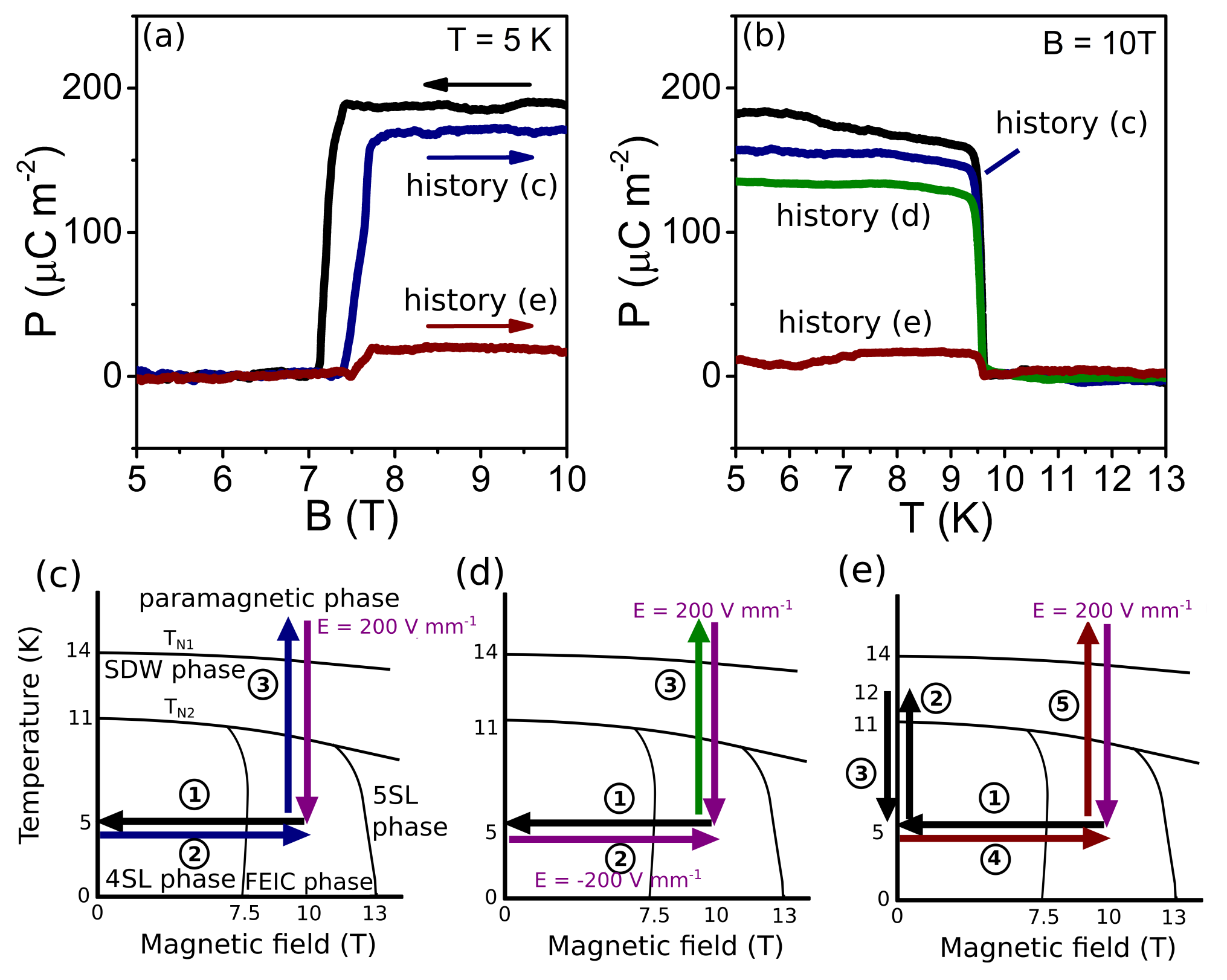}
\caption{Dependence of the polarization memory on the sample history. (a) Magnetocurrent and (b) pyrocurrent measurements (all taken in the absence of any bias $E$ field), showing the dependence of the polarization memory on three sample histories, as shown in (c)-(e). The schematic of the phase diagram is based on that reported by Kimura \textit{et al.} \cite{Kimura06}. In each case, there is an initial cool at \SI{10}{\tesla} with $E$ = \SI{200}{\volt\per\milli\metre} poling field, which gives the polarization shown by the black curves. (c) One leaves and returns to the FEIC phase with zero bias, recovering 88\% of $P$ (blue curves). (d) A reverse bias of \SI{-200}{\volt\per\milli\metre} is applied upon returning into the FEIC phase, which reduces $P$ to 73\% of its initial value, in the direction of the initial poling field (green curve). (e) A thermal cycle from 5 to 12 to \SI{5}{\kelvin} is performed at \SI{0}{\tesla}, which subsequently reduces $P$ to only 11\% of its initial value (red curves).}
\label{fig:memory}
\end{figure}

\begin{figure}[t]
\includegraphics[width=0.5\textwidth]{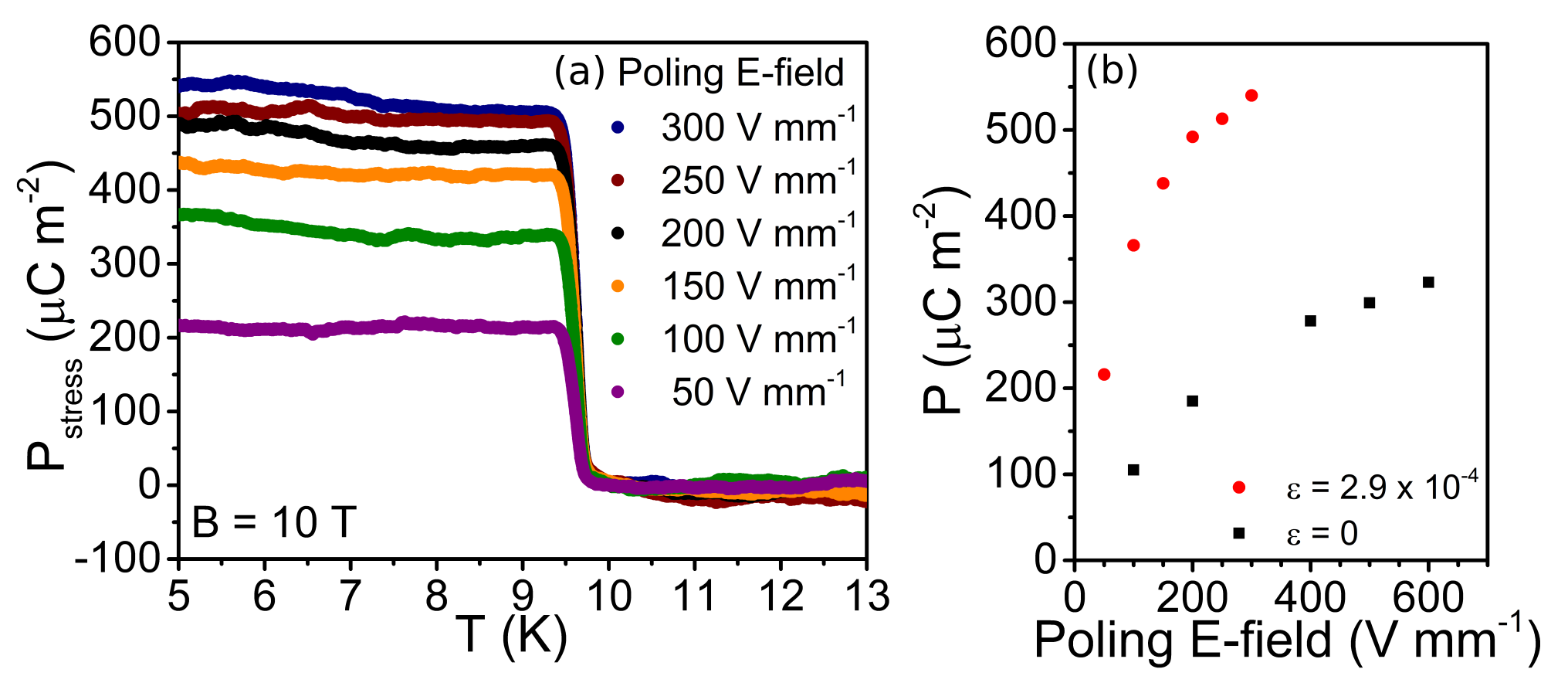}
\caption{(a) Bias dependence of the polarization under applied stress ($\sim$\SI{110}{\mega\pascal} at room temperature). (b) A significant polarization increase is seen under applied stress compared to the unstrained sample, and the polarization magnitude starts to saturate at lower $E$. This increase and $E$-field dependence is well explained by the formation of a single monoclinic $q$ domain, which increases the average $E$ experienced by the ferroelectric domains and causes a greater component of $P$ along [110] by removing symmetry-equivalent domains (see text). It is likely that there will also be some strain enhancement of polarization, as seen in CuFe$_{1-x}$Ga$_{x}$O$_{2}$ \cite{Mitsuda12}.
}
\label{fig:strain}
\end{figure}

\begin{figure}[t]
\includegraphics[width=0.5\textwidth]{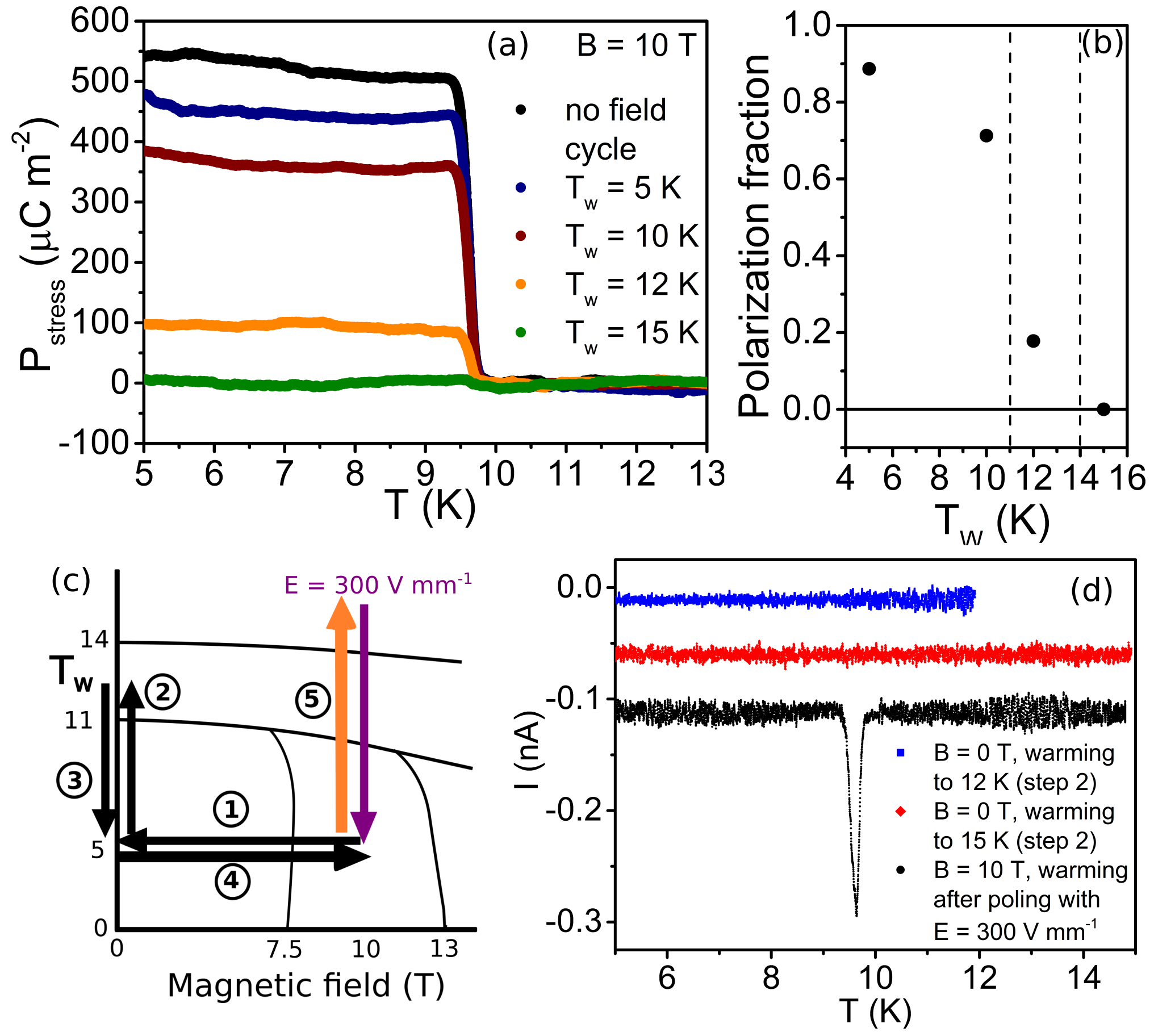}
\caption{Characterization of the memory effect under applied stress. 
(a,b) Dependence of the polarization memory on the sample history, for a history as illustrated in (c), in absolute values and as a polarization fraction. Dashed lines indicate the magnetic transition temperatures. In each measurement, after poling the sample and reducing $B$ to \SI{0}{\tesla}, a thermal cycle is performed up to a maximum warming temperature $T_{\text{W}}$, and the subsequent polarization compared to the initial polarization state (black curve). A large fraction (0.71) of the initial polarization is recovered for a thermal cycle to \SI{10}{\kelvin}. However the polarization memory is drastically reduced when the thermal cycle goes to \SI{12}{\kelvin} and exits the ground magnetic state, and a thermal cycle to \SI{15}{\kelvin} completely removes the memory effect, showing a strong correlation between the magnetic state and the memory effect.
(d) Pyrocurrents measured upon warming during the thermal cycle (each offset by \SI{-0.05}{\nano\ampere}) showed no evidence of any depolarization at the magnetic phase boundaries or thermally stimulated currents. The pyrocurrent from the ferroelectric phase at \SI{10}{\tesla} is shown for comparison.
}
\label{fig:strainmemory}
\end{figure}

Figure~\ref{fig:bias} shows the results of pyrocurrent measurements in a persistent magnetic field.
Figures~\ref{fig:bias}(a) and ~\ref{fig:bias}(b) show the polarization dependence on the poling $E$ field as a function of temperature at \SI{10}{\tesla}.
A strong poling-field dependence is observed, with a high rate of change of $P$ at low $E$ fields and saturation of $P$ starting to occur only for $E$ fields above \SI{400}{\volt\per\milli\metre}. Such a poling field dependence implies that at low $E$ fields there is a significant energy barrier for motion of the magnetic/ferroelectric domain walls that prevents the formation of a ferroelectric monodomain \cite{Nakajima214423}.
Nevertheless, the ferroelectric phase does show complete reversibility with the poling $E$-field direction when the system is warmed into the paramagnetic phase between subsequent measurements, as shown in Fig.~\ref{fig:bias}(c). Therefore there is no intrinsic coupling with $B$ that favors one ferroelectric domain type over the other, and no memory of the polarization direction after warming into the paramagnetic phase, as expected.

Having established absolute values of $P$ at $T$ = \SI{5}{\kelvin}, $B$ = \SI{10}{\tesla} through the pyrocurrent measurements [Fig.~\ref{fig:bias}(a)], we were then able to investigate whether there is any physical residual polarization in the 4SL phase through magnetocurrent measurements.
Figure~\ref{fig:bias}(d) shows the result of magnetocurrent measurements upon reducing $B$. $P$ was found to reduce precisely to zero within error at the magnetic phase transition at \SI{7.5}{\tesla}, giving no indication of a residual ferroelectric polarization. Furthermore, no depolarization of the 4SL phase was observed upon warming for $B$=\SI{0}{\tesla}. Therefore one can effectively ascribe the residual polarization observed in pulsed-field measurements to an artifact due to the applied bias, which must be applied throughout the measurement due to the short timewidth of the pulse.  One implication of this finding is that the crystal structure in the 4SL should be described by a nonpolar space group such as $P2/c$, rather than, for example, by $P2$.  Another implication is that the volume of residual FEIC regions within the 4SL phase, which has been proposed as a possible explanation of multiferroic memory effects, must be extremely small, or we would detect a pyroelectric current upon warming into the paramagnetic phase at $B$ = \SI{0}{\tesla}.  

We further investigated the polarization memory effect through a series of persistent-field measurements following different histories, as shown in Fig.~\ref{fig:memory}. 
Figures~\ref{fig:memory}(a),and ~\ref{fig:memory}(b) show the polarization as a function of $B$ and $T$, for three sample histories following a \SI{10}{\tesla} field-cool with a poling field $E$=\SI{200}{\volt\per\milli\metre}, as illustrated in Figs.~\ref{fig:memory}(c)-(e). First, the magnitude of the polarization memory was quantified by reducing $B$ to \SI{0}{\tesla} before reentering the ferroelectric phase with zero applied bias and then warming to the paramagnetic phase at a constant $B$ field of \SI{10}{\tesla} [see Fig~\ref{fig:memory}(c)]. The polarization recovered 88\% of its initial value, as shown by the magnetocurrent and pyrocurrent measurements (blue curves), compared with the initial polarization (black curves). Secondly, the polarization memory was found to be robust against a reversal of the electrical bias [see Fig~\ref{fig:memory}(d)]. When an $E$ field of \SI{-200}{\volt\per\milli\metre} is applied upon reentering the ferroelectric phase by increasing $B$, it is found that 73\% of the initial polarization is retained $\textit{in the direction of the initial poling E field}$ (green curve), a behaviour similar to the pulsed field measurements. Finally, the effect of warming was investigated [see Fig~\ref{fig:memory}(e)]. Having poled the sample while cooling at \SI{10}{\tesla} and then reduced the field to \SI{0}{\tesla}, a thermal cycle from 5 to 12 to \SI{5}{\kelvin} was performed. Then, upon reentering the ferroelectric phase, the polarization was found to be only 11\% of its initial value (red curves), indicating that the polarization memory had been almost but not completely removed in the SDW phase.

It has been previously proposed that the memory effect in CuFeO$_2$ may be associated with the presence of  magnetoelastic  $q$ domains, which  appear upon threefold symmetry breaking at $T_{\text{N1}}$, possibly through charging of the domain boundaries \cite{Mitamura07, Mitsuda2009, NakajimaJPSJ13}. In order to investigate this, we repeated the measurements for a crystal held under an applied uniaxial stress.  As described earlier, a stress was applied to the sample resulting in a strain of $\epsilon = 2.9 \times 10^{-4}$, which corresponds to a pressure of approximately \SI{110}{\mega\pascal}. Nakajima \textit{et al.} \cite{NakajimaJPSJ11} found that the application of \SI{10}{\mega\pascal} of uniaxial pressure on a [1$\overline{1}$0] surface producted a volume fraction of 0.973 of the (110) domain and it is known that the magnetic ground state is stable up to pressures above \SI{2}{\giga\pascal} \cite{Terada14}. Therefore we expect to form almost a complete magnetic monodomain upon cooling through $T_{\text{N1}}$, while maintaining the FEIC phase in applied magnetic fields.

Figure ~\ref{fig:strain}(a) shows the measured $P_{[110]}$ under applied stress, while Fig.~\ref{fig:strain}(b) shows a comparison of the poling-field dependence of $P_{[110]}$ in the presence and absence of uniaxial stress. As in the case of the unstrained sample, there is a significant poling-field dependence of $P$, but it is clear that $P$ is significantly higher in the strained samples.  This is completely consistent with the removal of the (1$\overline{2}$0) and ($\overline{2}$10) $q$ domains: in fact, as the whole crystal distorts along the [110] direction rather than along three symmetry equivalent directions, a larger component of $P$ along the [110] direction will be measured.  In the limiting case when $P$ is fully saturated within all monoclinic domains, one expects an increase by a factor of 1.5 upon straining. Taking $P_{\epsilon=0}$ = \SI{325}{\volt\per\milli\metre} for the limiting polarization of the unstrained sample, one would expect $P = $ \SI{488}{\micro\coulomb\per\square\metre} for the strained sample. In fact, $P_{\epsilon}$ is even higher, with a value of \SI{540}{\micro\coulomb\per\square\metre} at \SI{300}{\volt\per\milli\metre} and still increasing with $E$.  This is most likely due to the fact that, for a given poling $E$ field, in the unstrained sample the component of $E$ along the direction of $P$ is reduced in the (1$\overline{2}$0) and ($\overline{2}$10) directions, reducing the net $P$ within these domains, whereas the strained sample sees the full poling field in the $[110]$ direction.  Our polarization measurements confirm that we have an essentially monodomain sample below  $T_{\text{N1}}$. The additional increase in $P$ above the predicted factor of 1.5 could also be due in part to a strain enhancement of polarization, as seen in CuFe$_{1-x}$Ga$_{x}$O$_{2}$ \cite{Mitsuda12}, where the application of $\sim$ \SI{80}{\mega\pascal} caused a further increase in $P$ of $\sim$ 10\%.

To determine whether or not the memory effect is associated with charging at $q$-domain boundaries as suggested in previous studies, we investigated the correlation of the polarization memory with the magnetothermal history of the strained (monodomain) sample.
Figure~\ref{fig:strainmemory} shows the results of polarization measurements under uniaxial stress.
A series of measurements was performed where the strained sample was initially poled with $E$ = \SI{300}{\volt\per\milli\metre} upon cooling at $B$ = \SI{10}{\tesla} to $T$ = \SI{5}{\kelvin}. After removal of $E$ and short circuit of the electrodes, $B$ was reduced to \SI{0}{\tesla}, followed by a thermal cycle from \SI{5}{\kelvin} to $T_W$ to \SI{5}{\kelvin}, where $T_W$ is the maximum warming temperature reached. A set of values of $T_W$ = \SI{10}{\kelvin}, \SI{12}{\kelvin} and \SI{15}{\kelvin} was chosen to compare the effect of remaining in the 4SL phase, entering the SDW phase, and entering the paramagnetic phase. $B$ was then increased to \SI{10}{\tesla} before the polarization was measured upon warming at \SI{10}{\tesla} through a pyrocurrent measurement. This is compared to a measurement where no thermal cycle was performed ($T_W$ = \SI{5}{\kelvin}) and also to the initial polarization state (no field cycle; black curve). 

These data clearly demonstrate that the application of stress did not have a significant effect on the polarization memory. When no thermal cycle is performed ($T_W$ = \SI{5}{\kelvin}), $89\%$ of $P$ is recovered. Performing a thermal cycle to $T_W$ = \SI{10}{\kelvin} still maintained a significant polarization memory of $71\%$, however once the thermal cycle took the system out of the ground magnetic state the polarization memory was reduced to $18\%$ for $T_W$=\SI{12}{\kelvin} and was removed completely by warming into the paramagnetic state at \SI{15}{\kelvin}. Furthermore, there was no evidence of any decharging or depoling of the sample upon warming during the thermal cycles, as shown in Fig.~\ref{fig:strainmemory}(d). The polarization memory is not related to the presence of $q$-domain boundaries but is strongly related to the magnetic state of the material. The only other known type of \emph{in-plane} domain walls capable of becoming charged upon poling would be produced by obverse/reverse domains;  however, as discussed in Sec. II, our sample contains only a negligible amount ($<3\%$) of the alternate obverse/reverse domain.  On this basis, we can conclusively rule out  charging at \emph{structural} domain walls as an explanation of the observed memory effect in CuFeO$_2$ (see the Sec. IV for the possible role of \emph{magnetic} domain walls).

\begin{figure*}[t]
\includegraphics[width=1.0\textwidth]{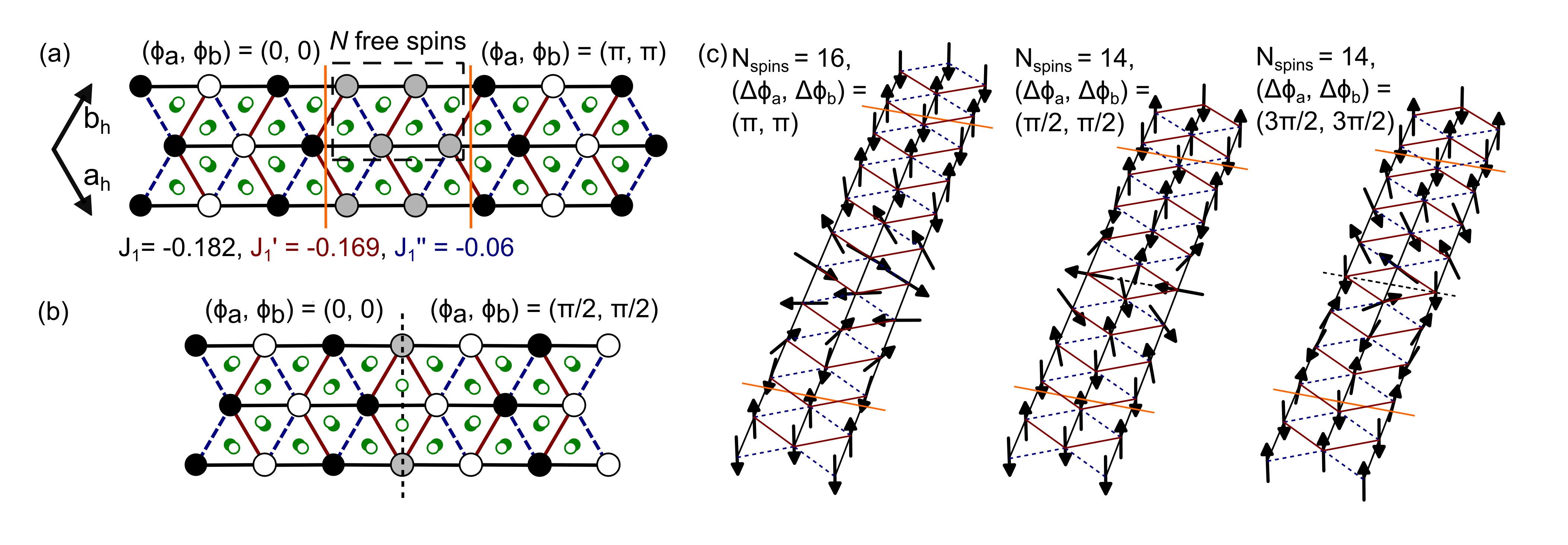}
\caption{(a) Bond ordering of the Fe atoms for antiferromagnetic phase domains with a phase shift of $\pi$ (black = $\downarrow$, white = $\uparrow$). Open green circles indicate the oxygen positions, which are displaced from their high-temperature positions (filled circles).
In the simulations, the outer spins were fixed to give the required phase difference, while the central $N$ spins (gray) were relaxed. Periodicity along the [1$\overline{1}$0] direction was imposed and previously reported exchange constants were used \cite{NakajimaPRB11}. (b) Phase boundaries of $\pi/2$ and $3\pi/2$ imply a break in the bond order, therefore we introduce a structural boundary (dashed line), where the bond order is reversed. (c) Minimum energy spin configurations for the three types of domain walls, calculated by a Monte Carlo method. The domain wall with a phase difference of $\pi$ was found to contain a full helix with $q\sim0.21$, which closely resembles the spin structure of the FEIC phase. Domain walls for phase differences of $\pi/2$ and $3\pi/2$ were found to contain a helix on one row of spins and antiferromagnetic order on the other. Therefore all phase domains allow the handedness of the ferroelectric phase to be maintained within the 4SL domain walls.}
\label{fig:4SLwalls}
\end{figure*}

\section{Discussion}

On the basis of the set of measurements presented in this paper, we can conclude that the previously observed residual polarization at $B$ = \SI{0}{\tesla} is an artifact, whereas  we observe a very robust ferroelectric memory effect, which is not strongly dependent on the structural domain composition of the crystal.  An important additional observation is that once the initial polarization state is set, the memory does not appear to be strongly sensitive to further electrical biasing, as seen in pulsed field switching measurements and upon application of a reverse bias in persistent-field measurements (Fig.~\ref{fig:memory}).

As previously discussed, it appears that charged domain boundaries are not likely to account for the observed memory effect. The enhanced polarization under strain suggests that the crystal has almost completely formed one monoclinic $q$ domain, and for $E$ = \SI{300}{\volt\per\milli\metre} under stress the ferroelectric domain population is expected to be close to 100\% in the direction of $E$.   Most importantly, the polarization memory appears to be largely unchanged by the application of strain, with $\sim 88\%$ polarization memory in both cases.  The fact that the polarization memory fraction is very similar in both the unstrained and the strained sample, even when the initial polarization state is significantly different (185 and \SI{540}{\micro\coulomb\per\square\metre} respectively) suggests that the ground state is, to a good extent, preserving the balance of helical domains in the poled FEIC phase. Additionally, the dependence on the magnetic history in Fig.~\ref{fig:strainmemory}, especially the fact that the polarization memory is drastically reduced in the SDW phase at 12 K and $B$ = \SI{0}{\tesla},  implies that the memory is strongly correlated with the magnetic configuration in the 4SL phase.

An explanation of the memory effect that is consistent with our data is related to the possibility of the 4SL and 5SL phases themselves retaining a non-bulk helicity through Bloch-type magnetic domain walls.  
Upon reentering the FEIC phase without bias or poling,  these domain walls would act as nucleation centers for the FEIC phase, producing a FEIC phase with a similar helicity configuration.

We further investigated this scenario by calculating the minimum energy classical spin configurations for different phase-slip domain walls in the 4SL phase using a Monte Carlo method. A three-dimensional spin Hamiltonian determined in a previous study by Nakajima \textit{et al.} \cite{NakajimaPRB11} was used, including in-plane next-nearest neighbour exchange interactions and axis anisotropies. 
Given that the ground state consists of four sublattices of spins, magnetic domain walls will occur with phase differences of $\pi/2$,  $\pi$, and $3\pi/2$, depending on where each domain nucleates on the lattice. As the bond ordering is a direct response to the magnetic ordering, phase differences of $\pi/2$ and $3\pi/2$ imply a break in the bond order, whereas $\pi$ domains imply that the bond order of a single-phase domain is maintained. Therefore for a $\pi$ phase boundary we consider spins on a lattice as shown in Fig.~\ref{fig:4SLwalls}(a), whereas for $\pi/2$ and $3\pi/2$ phase boundaries we introduce a structural domain boundary as shown in Fig.~\ref{fig:4SLwalls}(b). The two structural domains on either side of the boundary are the antiphase domains that would be formed due to loss of $C$ centering upon cooling through $T_{\text{N2}}$. These were not considered in our previous discussion but do not change the overall conclusions, since this type of domain wall does not produce any discontinuity in either the lattice or the atomic arrangement and cannot therefore accommodate additional charge.

In the simulations, the outermost spins were fixed to give the desired phase differences and the central $N$ spins were relaxed from a random initial orientation, with periodicity assumed in the [1$\overline{1}$0] direction. The classical minimum energy configurations for the three domain wall types are shown in Fig.~\ref{fig:4SLwalls}(c). For $\pi$ domains, the energy cost was found to be lowest for $N \geq 12$, with a total domain wall cost of \SI{1.49}{\milli\electronvolt}. The spins were found to form a helix in the plane perpendicular to the monoclinic $b$ axis with $q\sim0.21$, which closely resembles the spin structure of the ferroelectric phase. For $\pi/2$ and $3\pi/2$ domain walls, the lowest energy spin configurations consisted of one row of spins forming a helix, with the other row ordering antiferromagnetically, with some canting away from the $c$ axis. The total energy costs were 1.55 and \SI{2.65}{\milli\electronvolt} for the $\pi/2$ and $3\pi/2$ domain walls. 
Importantly, all three domain wall types have a \emph{definite helicity}, with an equal energy cost for structures with opposite helicities for a given phase boundary. 

This analysis enables us to propose the following scenario: during the phase transition from the FEIC to the 4SL phase, as domains of the 4SL phase nucleate and grow, the domain walls formed between them will maintain the helicity of the ferroelectric phase.  The energy cost of these walls is low compared to the thermal energy; $k_{B}T$ per spin is \SI{0.43}{\milli\electronvolt} at \SI{5}{\kelvin} compared to the average energy cost per spin in the domain wall of \SI{0.13}{\milli\electronvolt}, and so this should not significantly hinder the formation of magnetic domains.
By contrast, upon initially cooling in zero $B$ and $E$ field, there should exist an equal proportion of left- and right- handed domain walls, which could hinder the formation of a helical monodomain upon application of the $B$ field, as seen here as well as in previous studies \cite{Mitamura07}.
We remark that this concept is reminiscent to the nano-regions suggested for similar memory effects in CuO \cite{WuCuO2010} and MnWO$_4$ \cite{Taniguchi09}.  However, in this case the helical domain walls minimize the spin-configuration energy in the presence of antiferromagnetic domains of a \emph{single} magnetic phase, as opposed to being small metastable regions of the FEIC phase embedded within collinear order.

We now briefly discuss some implications of the helical-domain-wall model for the interpretation of our data.  The presence of helical domain walls naturally explains why the memory is not particularly sensitive to a reversing bias, as these domain walls do not give rise to a bulk polarization and have a high energy barrier to flip the handedness of rotation. However, upon warming at \SI{0}{\tesla} into the SDW phase, the domain structure would be severely rearranged, thus explaining the large reduction in memory for $T_W$ = \SI{12}{\kelvin}. The fact that there is still a small memory for $T_W$ = \SI{12}{\kelvin} may be due to phase coexistence of the 4SL and SDW phases, as no memory was observed for warming into the SDW phase at \SI{10}{\tesla}.
Furthermore, at the FEIC-5SL phase transition there is a change in bond ordering from the scalene-triangle order to a bond-ordered superlattice with $C$2/$m$ symmetry \cite{NakajimaPRB13}, as opposed to the 4SL-FEIC phase transition, where the bond ordering is maintained across the transition. Therefore a complete description of the magnetic bond ordering throughout the domain boundaries of the 5SL phase cannot readily be determined, preventing the extension of our calculations to the 5SL phase-domain structures.
Our experimental results reveal a weaker memory effect in the 5SL phase, which suggests that only some domain walls in the 5SL phase contain helical spin structures. 
Helical domain walls with the same magnetic structure as the polar FEIC phase would also be expected to be polar, with the same dipole moment per unit cell. However, the domain walls are not expected to give a macroscopically measurable polarization, as the measured compensating charge at the sample contacts is dependent on the volume fraction of the domain walls relative to the sample volume, which is expected to be negligible.

In summary, we have shown that a strong polarization memory exists in the nonpolar magnetic ground state of CuFeO$_2$ following electrical biasing of the ferroelectric phase at \SI{7.5}{\tesla} $<$ $B$ $<$ \SI{13}{\tesla}, which is largely insensitive to the magnetoelastic domain composition. This memory effect can be explained by the existence of helical domain walls within the ground state, which are found to be the lowest energy domain wall spin configurations in Monte Carlo simulations. More generally, given that similar memory effects are also reported for CuO \cite{WuCuO2010} and MnWO$_4$ \cite{Taniguchi09}, polarization memory effects caused by helical or cycloidal domain walls may be a universal feature of ferroelectric-incommensurate to collinear-antiferromagnetic  transitions. 
Clearly, memory in a non-ferroic phase through domain walls is a very interesting concept that could be developed further to extend the temperature and magnetic-field range of magnetoelectric coupling in multiferroic systems.

\acknowledgements

This work was funded by EPSRC Grant No. EP/J003557/1, entitled New Concepts in Multi- ferroics and Magnetoelectrics, and EPSRC Grant No. EP/M020517/1, entitled Oxford Quantum Materials Platform Grant. We would like to thank Dr. P. Manuel for assistance with and access to an Oxford Instruments High Field Magnet, Dr. P. Goddard for use of and assistance with the Nicholas Kurti Magnetic Field Laboratory, which is supported by the University of Oxford and the EPSRC, and J. Nutter for assistance in setting up experiments.  
R. D. J. acknowledges support from a Royal Society University Research Fellowship.
In accordance with the EPSRC policy framework on research data, access to the data will be made available from Ref. [\cite{CuFeO2data}].

\appendix*

\section{Details of crystal growth}
Polycrystalline CuFeO$_2$ powder was prepared in two steps using high purity ($>$ 99.99\%) CuO and Fe$_2$O$_3$ chemicals by the solid state reaction technique.  First, stoichiometric mixed chemicals were sintered in air at \SI{850}{\celsius} for \SI{64}{\hour} with several intermediate grindings to produce the precursor (CuFe$_2$O$_4$ +CuO).  In the second step, the precursor was sintered in a high purity argon atmosphere flow (\SI{35}{cc\per\minute}) at \SI{950}{\celsius} for \SI{64}{\hour} with intermediate grinding to get the phase pure CuFeO$_2$.  Using this powder, feed rods of \SI{8}{\milli\metre} diameter and \SI{12}{\centi\metre} long were prepared using a hydrostatic press and sintered again in high purity argon at \SI{1000}{\celsius} for \SI{24}{\hour}.

The crystals were grown using four-mirror optical floating-zone furnace (Crystal Systems Inc.).  The feed rod was scanned at a lower growth rate of 0.5-\SI{1}{\milli\metre\per\hour} with a counter-rotation of both feed and seed rods at 15-20 rpm in a high purity argon flow atmosphere (\SI{100}{\milli\liter\per\minute}). At the start of the growth, due to the peritectic reaction Cu$_2$O had formed as a second phase.  But, by carefully adjusting the lamp powder and Fe$_2$O$_3$ rich melt which act as a self-flux, we were able to produce a high quality single crystal with strong facets.


%

\end{document}